Article Subject: Spin Electronics

# Robustness and Scalability of p-bits Implemented with Low Energy Barrier Nanomagnets

Justine L. Drobitch[1,*], and Supriyo Bandyopadhyay[1,**]

[1] *Department of Electrical and Computer Engineering, Virginia Commonwealth University, Richmond, VA 23284, USA*
\* *Student Member, IEEE*
\*\* *Fellow, IEEE*

**Abstract**— Probabilistic (p-) bits implemented with low energy barrier nanomagnets (LBMs) have recently gained attention because they can be leveraged to perform some computational tasks very efficiently. Although more error-resilient than Boolean computing, p-bit based computing employing LBMs is, however, not completely immune to defects and device-to-device variations. In some tasks (e.g. binary stochastic neurons for machine learning and p-bits for population coding), extended defects, such as variation of the LBM thickness over a significant fraction of the surface, can impair functionality. In this paper, we have examined if unavoidable geometric device-to-device variations can have a significant effect on one of the most critical requirements for probabilistic computing, namely the ability to "program" probability with an external agent, such as a spin-polarized current injected into the LBM. We found that the programming ability is fortunately not lost due to reasonable device-to-device variations. The little variation in the probability versus current characteristic that reasonable device variability causes can be suppressed further by increasing the spin polarization of the current. This shows that probabilistic computing with LBMs is robust against small geometric variations, and hence will be "scalable" to a large number of p-bits.

## I. INTRODUCTION

Probabilistic p-bits are random bits which fluctuate between 0 and 1 [1]. They are neither deterministic bits used in classical Boolean logic, nor qubits which are coherent superposition of 0 and 1. Probabilistic computing with p-bits encoded in the magnetization states of low energy barrier nanomagnets (LBMs) is extremely energy-efficient and far more error-resilient than energy-efficient Boolean computing with nanomagnets, which is normally very error-prone [2]. Computing with p-bits has also been shown to excel in certain tasks such as combinatorial optimization [3], invertible logic [4] and integer factorization [5].

A popular platform for implementing p-bits is a low barrier nanomagnet (LBM) with two degenerate energy minima separated by a low energy barrier on the order of the thermal energy $k_BT$ ($k_B$ = Boltzmann constant and T = absolute temperature) [1]. In such a nanomagnet, the magnetization will fluctuate between the two orientations corresponding to the two degenerate energy minima because of thermal fluctuations. If we take a snapshot of the magnetization at any instant of time, it will point in a random direction. However, if its component along one of the two orientations is positive, then we will interpret the magnetization to represent the bit 1, while if it is negative, we will interpret it as bit 0. The bit will thus always fluctuate between 0 and 1 (sometimes 0 and sometimes 1) and act as a p-bit.

If the energy barrier is symmetric between the two degenerate minima, then bits 0 and 1 will be generated with equal probability. However, one can change that by passing a spin polarized current through the nanomagnet with spins polarized along one of the two orientations. This will bias the probability, either toward 0 or toward 1, depending on the current's magnitude and spin polarization (say, for example, 30% probability of 0 and 70% of 1 for a current of magnitude 1 μA with spins polarized in the direction representing bit 1). Such an approach provides a means to "program" the probability, which is the basis of probabilistic computing. It is also the basis of binary stochastic neurons frequently used in stochastic neural networks and machine learning.

The programmability (or "control") will be lost if the magnitude of the current needed for a particular probability distribution (e.g. 30% for 0 and 70% for 1) varies significantly from one nanomagnet to another because of small variations in the nanomagnet's lateral dimensions or thickness. This will be debilitating for probabilistic computing and, at best, limit the number of p-bits that can be harnessed to build a "p-circuit", thereby making p-bits suffer from similar limitations on scalability that afflict qubits. It is this problem that we study. In the past, we have shown that extended defects in an LBM (e.g. thickness variation over a significant fraction of the surface) will radically alter the auto-correlation function of the magnetization fluctuation in time [6] and the fluctuation rate [7], which will, respectively, affect applications in, say, binary stochastic neurons for machine learning [8] and population coding [9]. However, these are less serious than losing control over the probability because the latter is crucial to probabilistic computing. Therefore, it is critical to examine the effect of device-to-device variations caused by fabrication imperfections on the ability to control probability in LBMs.

To investigate this issue, we have carried out stochastic Landau-Lifshitz-Gilbert simulations to study magnetization fluctuations in an LBM (with in-plane magnetic anisotropy) at room temperature in the presence of a spin polarized current injected perpendicular to the plane of the LBM. These simulations allow us to generate the probability of bit 1 (encoded in the magnetization state of the LBM) as a function of the spin polarized current magnitude and polarization, and examine how this probability function varies with small variations in the nanomagnet's lateral dimensions and thickness. Our results show that the probability function is insensitive to reasonable variations. This is reassuring since it establishes that probabilistic computing with p-bits is not impaired by reasonable device-to-device variation and hence a large number of p-bits can be harnessed for p-circuits, meaning that p-bits are largely scalable.

## II. SIMULATIONS

We consider an elliptical cobalt nanomagnet of nominal thickness 6 nm, major axis 100 nm and minor axis 99.7 nm (Fig. 1). This nanomagnet has in-plane magnetic anisotropy and because it has very small eccentricity (nearly circular), the shape anisotropy energy barrier separating the two stable orientations along the major axis (easy axis) is only 2.45 $k_B T$ at room temperature. We follow the time evolution of the magnetization in this nanomagnet in the presence of thermal noise and a spin-polarized current injected perpendicular to plane with spin polarization along the major axis by solving the stochastic Landau-Lifshitz-Gilbert equation:

$$\frac{d\vec{m}(t)}{dt} = -\gamma \vec{m}(t) \times \vec{H}_{total}(t) + \alpha \left( \vec{m}(t) \times \frac{d\vec{m}(t)}{dt} \right)$$
$$+ a\vec{m}(t) \times \left( \frac{\eta \vec{I}_s(t)\mu_B}{qM_s\Omega} \times \vec{m}(t) \right) + b\frac{\eta \vec{I}_s(t)\mu_B}{qM_s\Omega} \times \vec{m}(t) \quad (1)$$

where

$$\hat{m}(t) = m_x(t)\hat{x} + m_y(t)\hat{y} + m_z(t)\hat{z} \quad \left[ m_x^2(t) + m_y^2(t) + m_z^2(t) = 1 \right]$$

$$\vec{H}_{total} = \vec{H}_{demag} + \vec{H}_{thermal}$$

$$\vec{H}_{demag} = -M_s N_{d-xx} m_x(t)\hat{x} - M_s N_{d-yy} m_y(t)\hat{y} - M_s N_{d-zz} m_z(t)\hat{z}$$

$$\vec{H}_{thermal} = \sqrt{\frac{2\alpha kT}{\gamma(1+\alpha^2)\mu_0 M_s \Omega(\Delta t)}} \left[ G_{(0,1)}^x(t)\hat{x} + G_{(0,1)}^y(t)\hat{y} + G_{(0,1)}^z(t)\hat{z} \right]$$

The last term in the right hand side of Equation (1) is the field-like spin transfer torque and the second to last term is the Slonczewski torque. The inclusion of the field like torque is necessary since the magnetization state of the nanomagnet will have to be read by a magneto-tunneling junction, which will result in a field-like torque. The coefficients $a$ and $b$ depend on device configurations and following [10], we will use the values $a=1$, $b=0.3$. Here $\hat{m}(t)$ is the time-varying magnetization vector in the nanomagnet normalized to unity, $m_x(t)$, $m_y(t)$ and $m_z(t)$ are its time-varying components along the x-, y- and z-axis, $\vec{H}_{demag}$ is the demagnetizing field in the soft layer due to shape anisotropy and $\vec{H}_{thermal}$ is the random magnetic field due to thermal noise [11]. The different parameters in Equation (1) are: $\gamma = 2\mu_B\mu_0/\hbar$ (gyromagnetic ratio), $\alpha$ is the Gilbert damping constant, $\mu_0$ is the magnetic permeability of free space, $M_s$ is the saturation magnetization of the magnetostrictive soft layer, $kT$ is the thermal energy, $\Omega$ is the volume of the nanomagnet given by $\Omega = (\pi/4)a_1a_2a_3$, $a_1$ = major axis, $a_2$ = minor axis, and $a_3$ = thickness, $\Delta t$ is the time step used in the simulation (0.1 ps), and $G_{(0,1)}^x(t)$, $G_{(0,1)}^x(t)$ and $G_{(0,1)}^x(t)$ are three uncorrelated Gaussians with zero mean and unit standard deviation [11]. The quantities $N_{d-xx}, N_{d-yy}, N_{d-zz}$ $\left[ N_{d-xx} + N_{d-yy} + N_{d-zz} = 1 \right]$ are calculated from the dimensions of the nanomagnet following the prescription of ref. [12]. We assume that the charge current injected into the nanomagnet is $\vec{I}_s(t)$ and that the spin polarization in the current is $\eta$. The spin current is given by $\eta\vec{I}_s(t) = \eta|\vec{I}_s(t)|\hat{z}$ where $\hat{z}$ is the unit vector along the major axis as shown in Fig. 1. The various parameters for the simulation are given in Table I.

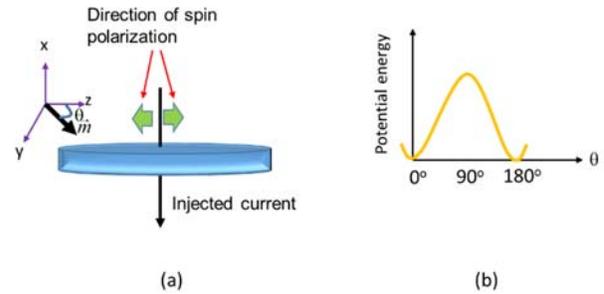

Fig. 1: (a) A slightly elliptical nanomagnet with in-plane magnetic anisotropy into which a spin-polarized current is injected perpendicular-to-plane. The spin polarization is along the major axis. The nanomagnet's dimensions are: major axis = 100 nm, minor axis = 99.7 nm and thickness = 6 nm. (b) The potential energy profile as a function of the in-plane magnetization orientation.

Stochastic Landau-Lifshitz-Gilbert simulations are run in the manner of ref. [11]. We start the simulation for any given magnitude and polarization of the spin polarized current with the initial value $m_x(0) = m_z(0) = 0$; $m_y(0) = 1$, i.e. the magnetization is initially pointing in one direction along the minor axis. We run the simulation for 10 ps and note the final value of $m_z$. If it is positive, then we interpret the magnetization state to represent the bit 1, while if it is negative, we interpret it as bit 0. One would measure the $m_z$ component with a magneto-tunneling junction (MTJ) whose hard layer is magnetized in one direction along the z-axis, and hence the resistance of the MTJ will be a measure the $m_z$ component. The resistance, of course, will not be binary and vary continuously between the high and low values since $m_z$ component will vary continuously between -1 and +1. Hence, a threshold function is used in probabilistic computing to interpret all positive $m_z$ component as bit 1 and all negative component as bit 0.

**Table I: Parameters used in the simulations**

| Parameters | Values |
|---|---|
| Saturation magnetization ($M_s$) | $1.1 \times 10^6$ A/m |
| Gilbert damping ($\alpha$) | 0.01 |
| Temperature (T) | 300 K |
| Spin polarization ($\eta$) | 0.3, 0.7 |
| Major axis (a1) | 100 nm |
| Minor axis (a2) | 90, 98, 99, 99.7 nm |
| Thickness (a3) | 5, 6, 7, 15 nm |

### III. RESULTS

We run 10,000 simulations of the magnetization dynamics for each value of spin polarized current (in steps of 0.1 mA) and calculate the fraction of simulations where the final state after 10 ps represents the bit 1. That fraction is the probability that the p-bit is 1 or $P(1)$. If we had monitored the bit as a function of time, this would have been the probability of observing the bit as 1, based on ergodicity. Obviously, $P(0)$ is always $1 - P(1)$, where $P(0)$ is the probability that the p-bit is 0. In Fig. 2, we show $P(1)$ as a function of the magnitude and spin polarization of the spin polarized current for four different nanomagnet thicknesses of 5 nm, 6 nm, 7 nm and 15 nm. Positive current corresponds to spin polarization along the +z-axis and negative current corresponds to polarization along the –z-axis. We plot the results for two different degrees of spin polarization $\eta$ in the current: 30% and 70%.

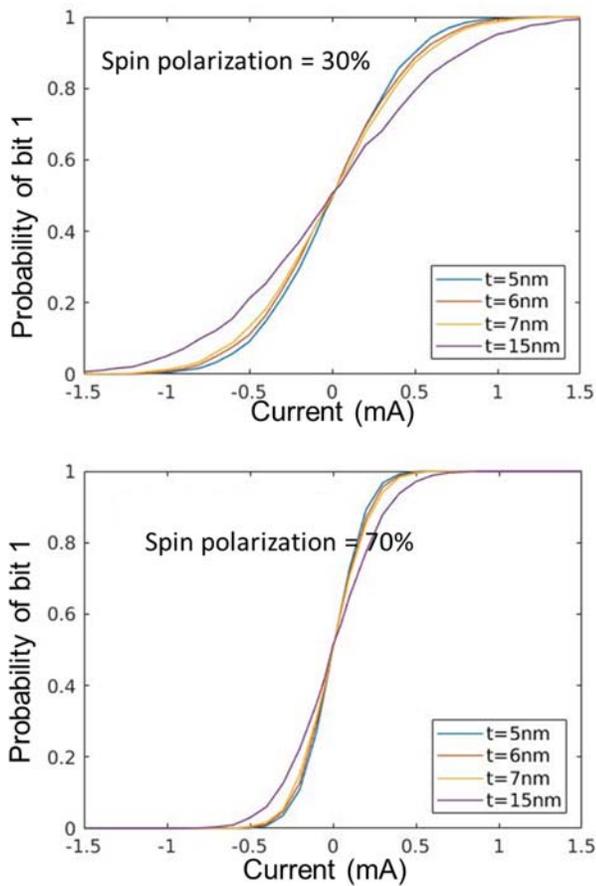

Fig. 2: The probability of bit 1 as a function of spin polarized current for four different nanomagnet thicknesses of 5, 6, 7 and 15 nm. The major axis is 100 nm and the minor axis is 99.7 nm. The results are plotted for two different degrees of spin polarization in the current: 30% and 70%. The variation in the probability at any given current is reduced at higher spin polarization. Positive value of the current corresponds to spin polarization in the +z-direction and negative values correspond to spin polarization in the –z-direction. For these thicknesses, the energy barrier heights are respectively 1.7 $k_BT$, 2.45 $k_BT$, 3.33 $k_BT$ and 15.29 $k_BT$

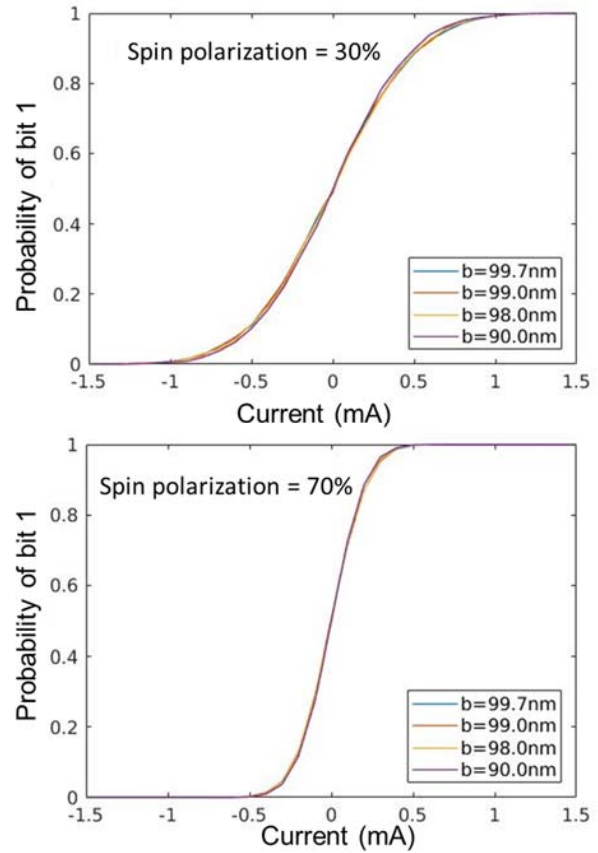

Fig. 3: The probability of bit 1 as a function of spin polarized current for four different nanomagnet minor axis dimensions of 90, 98, 99 and 99.7 nm. The major axis dimension is fixed at 100 nm and the thickness is 6 nm. The results are plotted for two different degrees of spin polarization in the current: 30% and 70%. As in Fig. 2, the variation is reduced at higher spin polarization.

In Fig. 3 we show $P(1)$ as a function of the magnitude and degree of spin polarization in the spin polarized current for four different minor axis dimensions of 99.7 nm, 99 nm, 98 nm and 90 nm (the major axis is fixed at 100 nm and the thickness is fixed at 6 nm). We keep the aspect ratio (major axis to minor axis dimension ratio) small enough so that the energy barrier in the nanomagnet remains sufficiently low (the energy barriers are 2.45 $k_BT$, 8.16 $k_BT$, 16.34 $k_BT$ and 81.62 $k_BT$ for the four different minor axis dimensions). When the minor axis is 90 nm, the energy barrier is obviously too high to qualify the nanomagnet as a "low barrier nanomagnet". Yet, even with that much variation in the barrier height (3300%), the probability curves change very little, showing that the probability versus current characteristic is very insensitive to barrier height variation in this regime. Comparing Figs. 2 and 3, we find that the probability versus current characteristic is much more insensitive to lateral dimension variation than thickness variation. This is fortunate since lateral dimension is more difficult to control since it is defined by lithography than thickness which is defined by film growth.

In Fig. 3, we show the plots for two different degrees of spin polarization $\eta$ in the current: 30% and 70%. The higher degree of spin polarization again suppresses the variability just as in the case of thickness variation.

The shapes of the probability curves in Figs. 2 and 3 are very similar to that in ref. [1], which explained the shape. At zero current, there will be no more preference for the p-bit to be 1 than to be 0. Hence, the probabilities of the p-bit being 0 and 1 will be equal and each will be exactly 0.5, like an unbiased coin with equal probability of head and tail. This is, of course, what we also observe. The current "biases" the probability toward either 0 or 1 depending on the direction of its spin polarization. At very high current levels, the probability of the p-bit being 1 will be nearly 100% for one sign of spin polarization and 0% for the other sign. Hence, $P(1)$ will saturate to 0 for very high current of one polarization and 1 for very high current of the opposite polarization. As a result, the probability $P(1)$ will have a dependence on the bias current of the type $P(1) \sim 0.5\tanh(I/I_0) + 0.5$, where $I$ is the spin-polarized current and $I_0$ is a constant, as discussed in ref. [1].

## IV. CONCLUSION

Clearly, the plots show that the probability curves are not affected much by *reasonable* variations in either thickness or lateral dimensions. In the case of thickness variation, we see a significant difference only for the 15 nm thickness. Variation in thickness by $\pm 1$ nm is reasonable since nanomagnets are usually fabricated on substrates with surface roughness of 0.3 nm, but the 15 nm thickness would require the thickness to vary by 9 nm from the target thickness of 6 nm and is not reasonable. Therefore, we can conclude that the probability curves are not affected significantly by reasonable thickness variations.

Variation in the lateral dimension (minor axis length) is even more forgiving. A variation of more than 9 nm, which is 9% of the minor axis dimension, does not make a significant difference in the probability curves. The little variation that there is can be further suppressed by increasing the degree of spin polarization in the current. All this means that the controlled probability generators are very reproducible and it is possible to make multiple generators with nominally identical characteristics because the device-to-device variation in the characteristics will be small. That also tells us that p-bits implemented with low energy barrier nanomagnets will be scalable up to a large number of p-bits because of the immunity to device-to-device variation.

Increasing the degree of spin polarization also decreases (expectedly) the magnitude of the current needed to pin the bit to either 0 or 1. All this is reassuring since it implies that the "control" over p-bits exercised with spin polarized current is *not* impaired by reasonable device-to-device variations and therefore a fairly large number of p-bits can be harnessed for "p-circuits" in many applications, i.e. p-bits are generally "scalable". This is in sharp contrast to qubits where only a small number can be entangled for quantum operations (the largest number entangled so far appears to be 53 [13]) because of decoherence. Classical p-bits do not suffer from decoherence and their scalability does not appear to be severely limited by reasonable device-to-device variations either. Some specific applications may still be vulnerable to defects [5, 6], but the practicality of implementing p-bits with LBMs is unassailable.

Finally, we clarify that the variations we have considered are uniform variations in thickness and lateral dimensions. Studying the effect of *spatially inhomogeneous* variations (e.g. surface roughness) would require running micromagnetic simulations instead of macrospin simulations (stochastic Landau-Lifshitz-Gilbert). Since running 10,000 micromagnetic simulations for each current value will be computationally prohibitive, we have not addressed the effect of spatially inhomogeneous variations (e.g. surface roughness) here. That will be addressed in a future work.

## ACKNOWLEDGMENT

This work was supported in part by the U.S. National Science Foundation under grants ECCS-1609303 and CCF-1815033.